\documentclass[preprint,prb,aps]{revtex4}
\usepackage{amsmath,amssymb}
\usepackage{graphicx}
\usepackage{bm}

\begin{document}

\title{On the construction of coherent states of position dependent mass Schr\"{o}dinger
equation endowed with effective potential}
\author{V~Chithiika Ruby and  M~Senthilvelan}
\affiliation{Centre for Nonlinear Dynamics, School of Physics,
Bharathidasan University, Tiruchirapalli - 620 024, India.}

\begin{abstract}
In this paper, we propose an algorithm to construct coherent states for 
an exactly solvable position dependent mass Schr\"{o}dinger equation.
We use point canonical transformation method and obtain ground state eigenfunction
of the position dependent mass Schr\"{o}dinger equation.  We fix the ladder operators in
the deformed form and obtain explicit expression of the deformed superpotential in terms
of mass distribution and its derivative.  We also prove that these deformed operators lead to
minimum uncertainty relations.  Further, we illustrate our algorithm with two examples
in which the coherent states given for the second example is new.
\end{abstract}
\pacs{03.65.Fd, 03.65.Ca, 03.65.Ge, 04.65.+e}

\maketitle
\section{\bf Introduction}
\label{sec1}
In a very recent paper Molski had presented a general scheme to construct minimum uncertainty coherent states for certain nonlinear oscillators\cite{mols}.  Coherent states \cite{Schr,Glau} are generally constructed by
(i) using the displacement operator technique or defining them as (ii) minimum uncertainty states
or (iii) annihilation operator eigenstates.  Even when such operators do not exist, different approaches\cite{Kla,How,Fer,Perel}
have been utilized to construct coherent states corresponding to different quantum mechanical
potentials\cite{Cruz,JU,midya}.  The scheme adopted by Molski is different from others in the sense that it adopts only a part of the basic concepts of the supersymmetric quantum
mechanics (SUSYQM).
In this work we extend Molski's scheme to the position dependent mass Schr\"{o}dinger equation (PDMSE) endowed with an effective potential and construct minimum
uncertainty coherent states for the PDMSE.

\par
The motivation to do this analysis comes from two reasons.  The primary reason comes from the
developments in the study of position dependent effective mass
Schr\"{o}dinger equation and the secondary reason arises from the renewed interest in the study of coherent states and its dynamics in nonlinear oscillators in recent times.  The contemporary studies
have shown that in a wide
variety of physical problems an effective mass depending on the position is of utmost
relevance.  To name a few problems, we cite (i) effective interactions in nuclear physics \cite{nuclear},
(ii) carriers and impurities in crystals \cite{crystal}, (iii) quantum dots \cite{qdot}, (iv) quantum liquids \cite{qliquid}, (v) semiconductor heterostructures \cite{hetero} and (vi) in neutron stars \cite{neutron}.  Apart from the experimental analysis various theoretical aspects of PDMSE
have also been investigated in detail, for example, exact solvability \cite{sol}, shape invariance \cite{shape},
quasi-exact solvability \cite{quasi}, supersymmetric or intertwining formulation \cite{intertwin}, Lie-algebraic approach \cite{lie} and
Green's approach \cite{green} have been studied widely.  We also note that in the above the quasi-exactly solvability, supersymmetric formulation and
Lie-algebraic approach are intimately related via nonlinear supersymmetry (see for example Ref. \onlinecite{ply} and references therein). 

\par
To construct the coherent states
of the PDMSE one needs to know the ground state wavefunction of it.  To obtain eigenfunction and energy values
of the PDMSE we use point canonical transformation method being followed in the literature.
In the second step one has to build suitable ladder operators.
Since Molski adopted a part
of basic concepts from the SUSYQM we look for the supersymmetric factorization
technique for PDMSE.  Interestingly, this has been worked out recently by Suzko and Schulze-Halberg in a
different context \cite{suz} (see also Ref. \onlinecite{intertwin} and references therein
for supersymmetric factorization of PDMSE).
Through supersymmetric factorization technique the authors have
described a method to construct the intertwining operators for the PDMSE.  However, after a detailed analysis,
we find that the intertwining operators considered by them
of the form $\hat A=a(x)\frac{d}{dx}+\phi(x)$ are not suitable for our present purpose.  While looking for a
more suitable form of intertwining operators
we find that the deformed momentum operators used by Quesne and Tkachuk \cite{ques} of the form,
$\pi = \sqrt{f(x)}p\sqrt{f(x)}$, where the function $f$ depends on the coordinates, are more suitable
to construct coherent states since such deformed operators lead to nonzero minimal uncertainties
in the position and momenta.
Based upon this observation we fixed the intertwining operator $\hat A$ of the
form $\hat A=a(x)\displaystyle{\frac{d}{dx}}a(x)+\phi(x)$, with unknown functions $a(x)$ and $\phi(x)$.  Once a suitable form of $\hat A$ has been chosen the rest of the work confined to find the explicit form of the functions $a(x)$ and $\phi(x)$.
We then follow the work of Suzko and Schulze-Halberg \cite{suz} and determine the functions
$a(x)$ and $\phi(x)$ explicitly, which comes out in terms of mass distribution $m(x)$ and its derivatives.  From the known expressions of $a(x)$ and $\phi(x)$ we build the operators  $\hat{A}$, $\hat{A}^{\dagger}$ and the deformed momentum.
We prove that the coherent states minimize the generalized position-momentum
uncertainty relation.  In this way we overcome the technical
difficulties and develop a general scheme to construct the minimum uncertainty coherent states for
an exactly solvable PDMSE.

We illustrate our scheme with two examples, namely (i) linear harmonic oscillator
and (ii) a new exactly solvable nonlinear oscillator which was introduced by Cari\~{n}ena {\it et al}  \cite{car1}.
We construct coherent states for the PDMSE associated with these two oscillators by considering
two different kinds of mass distributions which are often being used in semiconductor physics.
The energy values and eigenfunctions associated with the PDMSE associated with the harmonic oscillator
is already known.  In a
recent paper, Biswas and Roy \cite{Roy}, have constructed coherent states
exclusively for the effective mass harmonic oscillator through
displacement operator method.  Regarding the second example is concerned the solvability of PDMSE is discussed
only very recently by one of the
present authors \cite{sen}.  However, the coherent state for this PDMSE is being discussed for the
first time in this paper. Thus in a sense the methodology as well as the results given for an example are new to the literature.  
For the sake of comparison we also construct coherent states for the position dependent mass problem
using Perelomov's definition \cite{Perel}.  We find that the results obtained through these two different approaches agree with each other.

We organize our paper as follows.  In the following section, we briefly describe the method of
solving the PDMSE through point canonical transformation method.  In Sec. III, through intertwining
method we
construct suitable creation and annihilation operators for the PDMSE.  In Sec. IV, we prove that the states minimize the generalized position-momentum uncertainty
relation.  In Sec. V, we construct coherent states for two exactly solvable potentials associated with
a PDMSE.  In Sec VI, we briefly discuss Perelomov's approach and construct 
coherent states for the position dependent mass problem.  Finally, we present our conclusions in Sec. VII.

\section{\bf PDMSE and the methd of solving it}
\label{sec2}
In this section, we recall briefly the point canonical transformation approach to solve the PDMSE \cite{Alhai,Akt}.
In the case of the symmetric ordering of the momentum and mass the one dimensional
Schr\"{o}dinger equation with position dependent mass is given by
\begin{eqnarray}
-\frac{1}{2}\frac{d}{dx}\left[\frac{1}{m(x)}\frac{d{\tilde \psi_n}(x)}{dx}
\right]+{\tilde V}(x){\tilde \psi_n}(x) = {\tilde E_n} {\tilde \psi_n}(x).
\label{pct4}
\end{eqnarray}
Eq.~(\ref{pct4}) can be rewritten explicitly in the form
\begin{eqnarray}
\left(-\frac{1}{2}\frac{d^2}{dx^2}+\frac{m'}{2m}\frac{d}{dx}+m{\tilde V}(x)
\right){\tilde \psi_n}(x) = m{\tilde E_n} {\tilde \psi_n}(x).
\label{pct5}
\end{eqnarray}

One way of solving Eq.(\ref{pct5}) is to relate the latter with the
one dimensional time independent Schr\"{o}dinger equation with a constant mass,
\begin{eqnarray}
-\frac{1}{2}\frac{d^2\psi_n}{dy^2}+V(y)\psi_n(y) = E_n\psi_n(y),
\label{pct1}
\end{eqnarray}
where we have taken $\hbar=1$ and the mass $m=1$.  This can be done by introducing a transformation
$y \rightarrow x$ through a mapping function $y=f(x)$ and
\begin{eqnarray}
\psi_n(y) = g(x)\tilde{\psi_n}(x),
\label{pct2}
\end{eqnarray}
in the constant mass Schr\"{o}dinger equation so that the latter becomes
\begin{eqnarray}
\left[-\frac{1}{2}\frac{d^2}{dx^2}-\left(\frac{g'}{g}-\frac{f''}{2f'}\right)\frac{d}{dx}
-\frac{1}{2}\left(\frac{g''}{g}-\frac{g'f''}{gf'}\right)+(f')^2V(f(x))
\right]{\tilde \psi_n}(x) = (f')^2E_n {\tilde \psi_n}(x),
\label{pct3}
\end{eqnarray}
where prime denotes differentiation with respect to $x$.

Now comparing Eqs. (\ref{pct3}) and (\ref{pct5}) one observes that Eq. (\ref{pct3}) is
identical with Eq. (\ref{pct5})
if the following conditions satisfy \cite{Alhai,Akt},
\begin{eqnarray}
g(x) & =  m^{-\frac{1}{4}}(x), \quad f(x)  =  \int^x m^{\frac{1}{2}}(x')dx', \qquad m(x) > 0
\nonumber\\
{\tilde V}(x)  & =  V(f(x))+\displaystyle{\frac{1}{8m}
\left[\frac{m''}{m}-\frac{7}{4}\left(\frac{m'}{m}\right)^2\right]},  \quad
\tilde{E}_n  =  E_n.
\label{pct7b}
\end{eqnarray}
Here, $\tilde V$ is known as effective potential.  If we know the bound state energy
spectrum and the eigenfunctions of the reference potential, $E_n$ and $\psi_n(y)$ respectively,
 we can now construct the energy spectrum and eigenfunctions, $\tilde{E}_n$ and
$\tilde{\psi_n}(x)$ respectively, of the target potential $\tilde{V}(x)$ from Eq.(\ref{pct7b}), that is
\begin{eqnarray}
&\tilde{E}_n = E_n, \nonumber\\
&\tilde{\psi}_n(x) = \frac{1}{g(x)}\tilde{\psi_n}(y) = m^{\frac{1}{4}}(x)\psi_n(f(x)).
\label{pct6}
\end{eqnarray}
For a given mass distribution, $m(x)$, one can get
eigenvalues and eigenfunctions without solving the PDMSE (\ref{pct5})
by using the relations (\ref{pct7b}) and (\ref{pct6}).
The exactly solvable PDMSE is of the form
\begin{eqnarray}
\left(-\frac{1}{2}\frac{d}{dx}\left[\frac{1}{m(x)}\frac{d}{dx}
\right]+V(f(x))+\frac{1}{8m}
\left[\frac{m''}{m}-\frac{7}{4}\left(\frac{m'}{m}\right)^2\right]\right)
{\tilde \psi_n}(x) = {\tilde E_n} {\tilde \psi_n}(x).
\label{pct8}
\end{eqnarray}

In order to construct the coherent states for the potential $\tilde{V}$, we need to find the
ground state solution $|0\rangle$ of Eq. (\ref{pct8}) which is an eigenstate of the operator
$\hat{A}$.  If $\hat{A}$ annihilates the ground state
$\hat{A}|0\rangle=0$, then the coherent states $|\alpha\rangle$ are the eigenstates of the
annihilation operator $\hat{A}$ and the following relations are fulfilled \cite{mols}:
\begin{eqnarray}
\hat{A}|\alpha\rangle  = \alpha|\alpha\rangle,\;\;\;
\langle\alpha|\alpha^* = \langle\alpha|\hat{A}^{\dagger},\;\;\;
|\alpha\rangle  = |0\rangle\exp[{\sqrt{2}\alpha f(x)}].
\label{j31a}
\end{eqnarray}

{\it {The last relation in (\ref{j31a})
differs from the constant mass case}}.  The ground state wave function
$|0\rangle$ appearing in (\ref{j31a}) can be obtained by using the procedure given above.
Once we know the ground state solution then it can be utilized to construct the coherent states of the PDMSE.

\section{\bf Intertwinning technique}
\label{sec3}
Let us consider the effective mass Schr\"{o}dinger equation and its associated Hamiltonian $H^{(1)}$ in atomic
units
\begin{eqnarray}
H^{(1)}\tilde{\psi}^{(1)}=\tilde{E}^{(1)}\tilde{\psi}^{(1)}, \quad H^{(1)}=\left[-\frac{1}{2}\frac{d}{dx}\left(\frac{1}{m(x)}\frac{d}{dx}\right)+\tilde{V}_1(x)\right],
\label{j5}
\end{eqnarray}
where the energy $\tilde{E}^{(1)}$ is a real constant, $m(x)$ stands for position dependent mass, $\tilde{V}_1$ denotes
the potential and $\tilde{\psi}^{(1)}$ is the wavefunction.  Let us try to relate the problem (\ref{j5}) to a
problem of the same form but for a different potential, that is
\begin{eqnarray}
H^{(2)}\tilde{\psi}^{(2)} = \tilde{E}^{(2)}\tilde{\psi}^{(2)}, \quad
H^{(2)}=\left[-\frac {1}{2}\frac {d}{dx}\left(\frac{1}{m(x)}\frac{d}{dx}\right) + \tilde{V}_2(x)\right].
\label{j6}
\end{eqnarray}
In general, we have $\tilde{V}_2\neq \tilde{V}_1$ and $\tilde{\psi}^{(2)} \neq \tilde{\psi}^{(1)}$.  Now we
connect the problems (\ref{j5}) and (\ref{j6}) by means of the intertwining method.  To do this
we look for an operator $\hat A$ that satisfies the following intertwining relation, 
\begin{eqnarray}
\hat{A}H^{(1)} = H^{(2)} \hat A.
\label{j7}
\end{eqnarray}
The operator $\hat A$ is called intertwining operator for the Hamiltonians $H^{(1)}$ and $H^{(2)}$.  If
this intertwining relation is fulfilled then the solutions $\tilde{\psi}^{(1)}$ and $\tilde{\psi}^{(2)}$ are related
via
\begin{eqnarray}
\tilde{\psi}^{(2)}(x) = \hat A \tilde{\psi}^{(1)}(x).
\label{j7a}
\end{eqnarray}

We seek the operator, $\hat A$, of the form
\begin{eqnarray}
\hat{A}=\frac{1}{\sqrt{2}}\left[a(x)\frac{d}{dx}a(x)+\phi(x)\right],
\label{j8}
\end{eqnarray}
where $a$ and $\phi$ are to be determined such that $\hat{A}$ fulfills (\ref{j7}).  We note that the form of the intertwiner (\ref{j8}) is different from the one considered by Suzko and
Schulze-Halberg\cite{suz}.

To determine $a$ and $\phi$ we substitute the explicit form of the Hamiltonians (\ref{j5}) and (\ref{j6})
into the interwining relation (\ref{j7}) and allows it to operate on a function $\psi(x)$, that is
\begin{eqnarray}
\frac{1}{\sqrt{2}}\left[a\frac{d}{dx}a+\phi\right]\left[-\frac{1}{2}\frac{d}{dx}
\left(\frac{1}{m(x)}\frac{d}{dx}\right)+\tilde{V}_1(x)\right]\psi(x)
\nonumber \\
\qquad = \frac{1}{\sqrt{2}}\left[-\frac{1}{2}\frac{d}{dx}\left(\frac{1}{m(x)}\frac{d}{dx}\right)
+\tilde{V}_2(x)\right]\left[a\frac{d}{dx}a+\phi\right]\psi(x).
\label{j9}
\end{eqnarray}
Equating the coefficients of different derivatives of $\frac{d^k\psi}{dx^k},\;k=0,1,2,3$, to
zero, we get the following relations (after simplification)
\begin{eqnarray}
&\displaystyle{\frac{a'}{a}+\frac{m'}{4m}} = 0,
\label{j10}\\
& \tilde{V}_1-\tilde{V}_2 -\displaystyle{\frac{a'm'}{am^2}+\frac{m''}{2m^2}-\frac{m'^2}{m^3}
+\frac{2a''}{am}+\frac{2a'^2}{a^2m}+\frac{\phi'}{a^2m}}=0,
\label{j11}\\
& a^2\tilde{V}'_1 + (\tilde{V}_1-\tilde{V}_2)\left(\phi+aa'\right)+\displaystyle{\frac{1}{2m}(\phi''+3a'a''+aa''')
-\frac{m'}{2m^2}\left(\phi'+a'^2+aa''\right)}=0,
\label{j12}
\end{eqnarray}
where prime denotes differentiation with respect to $x$.

Integrating Eq. (\ref{j10}), we obtain
\begin{eqnarray}
a(x)=\frac{C}{m^{1/4}},
\label{j13}
\end{eqnarray}
where $C$ is an integration constant.  However, without loss of generality, one can set this
integration constant to one as this constant can be absorbed into the normalization constant of the
wavefunction.  Substituting Eq. (\ref{j13}) into Eq. (\ref{j11}) and simplifying the resultant
equation we arrive at
\begin{eqnarray}
\tilde{V}_2 = \tilde{V}_1 +\frac{\phi'}{\sqrt{m}}.
\label{j14}
\end{eqnarray}

With the definition of  Eq. (\ref{j14}), Eq. (\ref{j12}) reads
\begin{eqnarray}
a^2\tilde{V}'_1-\frac{\phi'}{\sqrt{m}}\left(\phi+aa'\right)-\frac{m'}{2m^2}\left(\phi'+a'^2+aa''\right)
+\frac{1}{2m}\left(\phi''+3a'a''+aa'''\right)=0.
\label{j15}
\end{eqnarray}

To integrate this equation let us introduce a transformation from $\phi$ to a new function $K(x)$
of the form $\phi=Ka^2-aa'$.  Substituting the expressions $a$ and $\phi$ into (\ref{j15}) and
simplifying the resultant equation we arrive at
\begin{eqnarray}
\tilde{V}'_1+\frac{K}{2}\left(\frac{1}{m}\right)''+\frac{K^2m'}{2m^2}-\frac{KK'}{m}+\frac{K''}{2m}
-\frac{m'K'}{m^2}=0.
\label{j16}
\end{eqnarray}
Eq. (\ref{j16}) can be rewritten as a perfect derivative of the form
\begin{eqnarray}
\frac{d}{dx}\left[\frac{1}{2m}\left(K^2-K'-v\right)-\frac{K}{2}\left(\frac{1}{m}\right)'\right]=0,
\label{j17}
\end{eqnarray}
where we have defined $\tilde{V}_1=\displaystyle{\frac{v}{2m}}$.  Upon integrating Eq. (\ref{j17}), one gets Riccati equation
of the form
\begin{eqnarray}
\frac{1}{2m}\left(K^2-K'-v\right)-\frac{K}{2}\left(\frac{1}{m}\right)'= - \lambda,
\label{j18}
\end{eqnarray}
where $\lambda$ is an integration constant.

This Riccati equation can be linearised through the transformation
\begin{eqnarray}
K = -\frac{u'}{u},
\label{j19}
\end{eqnarray}
where $u=u(x)$.  In the new variable, $u$, Eq. (\ref{j18}) reads
\begin{eqnarray}
-\frac{1}{2m}u''-\frac{1}{2}\left(\frac{1}{m}\right)'u'+Vu=\lambda u.
\label{j20}
\end{eqnarray}
Eq. (\ref{j20}) is nothing but the initial Eq. (\ref{j5}) at $\tilde{E} = \lambda $. Once
we know the solution $u=u(x)$ from (\ref{j20}) then we can fix $\phi$ be of the form
\begin{eqnarray}
\phi = Ka^2-aa' = \frac{K}{\sqrt{m}}+\frac{m'}{4m^{3/2}}.
\label{j21}
\end{eqnarray}
To construct $K$ and consequently $\phi$, we consider ground state solution.  With the definition of $a$ and 
$\phi$ (vide Eqs. (\ref{j13}) and (\ref{j21}) respectively) the operator $\hat{A}$ can be written
explicitly in the form
\begin{eqnarray}
\hat{A}=\frac{1}{\sqrt{2}}\left[m^{-1/4} \frac{d}{dx} m^{-1/4} + \phi\right].
\label{j22}
\end{eqnarray}
Correspondingly the creation operator can be written as
\begin{eqnarray}
\hat{A}^\dagger =\frac{1}{\sqrt{2}}\left[-m^{-1/4} \frac{d}{dx} m^{-1/4} + \phi\right],
\label{j23}
\end{eqnarray}
where $-im^{-1/4}\frac{d}{dx}m^{-1/4}$ is nothing but the deformed momentum ($\Pi$).
We note here that
{\it $\phi$ is called deformed superpotential} \cite{ques}.  For
more details about the deformed operators and their algebra one may refer the very recent paper
of Quesne \cite{ques2} and references therein.

Now we prove that the creation and annihilation operators given above reproduces the PDMSE in the factorized form.  
Using Eqs. (\ref{j22}) and (\ref{j23}), one can get 
\begin{eqnarray}
\hat{A}^\dagger \hat{A} = -\frac{1}{2}\frac{d}{dx}\left(\frac{1}{m(x)}\frac{d}{dx}\right)+\left[\frac{\phi^2}{2}-\frac{\phi^{'}}{2\sqrt{m(x)}}+\frac{m^{''}}{8m^2}-\frac{7}{32}\frac{m'^2}{m^3}\right]. 
\label{j23a}
\end{eqnarray}
Substituting Eq. (\ref{j21}) and its derivative in Eq. (\ref{j23a}) the latter simplifies to 
\begin{eqnarray}
\hat{A}^\dagger \hat{A} = -\frac{1}{2}\frac{d}{dx}\left(\frac{1}{m(x)}\frac{d}{dx}\right)+\frac{1}{2m(x)}\left[K^{2}+\frac{Km^{'}}{m}-K^{'}\right]. 
\label{j23b}
\end{eqnarray}
With the help of (\ref{j18}), Eq. (\ref{j23b}) can be further simplified to 
\begin{eqnarray}
\hat{A}^\dagger \hat{A} = -\frac{1}{2}\frac{d}{dx}\left(\frac{1}{m(x)}\frac{d}{dx}\right)+ \tilde{V}_1(x) - \lambda. 
\label{j23c}
\end{eqnarray}
If $\lambda \neq 0$ then the annihilation and creation operators given in Eqs. (\ref{j22}) and (\ref{j23}) respectively
reproduces the Hamiltonian $H$ ($H = H^{(1)}-\lambda$) in the factorized form, that is 
\begin{eqnarray}
H = \hat{A}^\dagger \hat{A}.
\label{j23d}
\end{eqnarray}
Hence, the solution satisfying Eq. (\ref{j31a}) is the coherent states
for the PDMSE corresponding to the Hamiltonian $H$.

The operator $\hat{A}$ cannot be used to generate solutions of (\ref{j6}) at energy $\lambda$,
since $\hat{A}u=0$.  It is known that the second order ordinary differential equation (\ref{j20})
admits two independent
solutions at any fixed value of $\lambda$.  Hence if we know one of them the other solution can be found
for the same eigenvalue.  Following the procedure given in Ref. \onlinecite{suz} one can construct the second
linearly independent solution of the form
\begin{eqnarray}
\tilde u = u \exp{\int {dx'\frac{m(x')}{|u(x')|^2}}},
\label{j24}
\end{eqnarray}
where $\tilde{u}$ is the second independent solution.  One can check that the action of $\hat{A}$
on the function $\tilde{u}$ gives us a solution $\eta$ of the transformed Eq. (\ref{j6}) at energy
$\lambda$, that is
\begin{eqnarray}
\eta = \hat{A}\tilde u  = \frac{\sqrt{m}}{u}.
\label{j25}
\end{eqnarray}

Finally, we express the operators $\hat{A}$ and $\hat{A}^{\dagger}$, (\ref{j22}) and (\ref{j23})
respectively, in terms of the function $\eta$ which are solutions
of (\ref{j6}) at the value $\lambda$.  To do this we rewrite $K$ in terms of $\eta$ by using the
relation
\begin{eqnarray}
K = -\frac{u'}{u} = \frac{\eta'}{\eta}-\frac{m'}{2m}=\tilde K-\frac{m'}{2m},\qquad \tilde K = \frac{\eta'}{\eta}.
\label{j27}
\end{eqnarray}

Hence, the deformed superpotential $\tilde \phi$ and the operators are written as
\begin{eqnarray}
\tilde{\phi} = \frac{\tilde K}{\sqrt{m}}-\frac{m'}{4m^{3/2}},
\label{j28}
\end{eqnarray}
\begin{eqnarray}
\hat{A}=\frac{1}{\sqrt{2}}\left[m^{-1/4} \frac{d}{dx} m^{-1/4} + \tilde{\phi}\right],\;\;\;
\label{j29}
\hat{A}^\dagger=\frac{1}{\sqrt{2}}\left[-m^{-1/4} \frac{d}{dx} m^{-1/4} + \tilde{\phi}\right].
\label{j30}
\end{eqnarray}
Hence, if we know the solution of the Eq. (\ref{j6}) at an energy $\lambda$, we can find out the
deformed superpotential, annihilation and creation operators.

\section{\bf Minimum uncertainty}
\label {sec4}
In the following we prove that the states $|\alpha\rangle$ minimize the deformed position -
momentum uncertainty relation\cite{mols}
\begin{eqnarray}
\left(\Delta \phi\right)^2.\left(\Delta \Pi \right)^2 \ge -\frac{1}{4} \langle\alpha|\left[\phi, \Pi\right]|\alpha\rangle^2,
\label{j31}
\end{eqnarray}
where,
\begin{eqnarray}
\left[\phi, \Pi\right] = i\frac{\phi'}{\sqrt{m}}.
\end{eqnarray}

To prove this relation, (\ref{j31}), let us calculate the uncertainties in the deformed momentum and superpotential, that
is
\begin{eqnarray}
\left(\Delta \phi\right)^2 = \langle\alpha|\phi^2|\alpha\rangle
-\langle\alpha|\phi|\alpha\rangle^2,\;\;\;
\left(\Delta \Pi \right)^2 =  \langle\alpha|\Pi^2|\alpha\rangle
-\langle\alpha|\Pi|\alpha\rangle^2.
\label{j34}
\end{eqnarray}
To evaluate the above expressions it is more convenient to express
the deformed momentum and deformed superpotential
in terms of operators $\hat{A}$ and $\hat{A}^{\dagger}$ (vide Eqs. (\ref{j22}) and (\ref{j23})), that is
\begin{eqnarray}
\phi = \frac{1}{\sqrt{2}}\left(\hat{A}+\hat{A}^\dagger\right),\;\;\;
\Pi = \frac{-i}{\sqrt{2}}\left(\hat{A}-\hat{A}^\dagger\right).
\label{j37}
\end{eqnarray}
The commutation relation between the two abstract operators can be evaluated to give
\begin{eqnarray}
\left[\hat{A},\hat{A}^\dagger\right] = \frac{\phi'}{\sqrt{m}}.
\label{j38}
\end{eqnarray}
With the help of (\ref{j38}) we find
\begin{eqnarray}
\langle\alpha|\phi^2|\alpha\rangle = & \frac{1}{2}(\alpha+\alpha^*)^2+\frac{1}{2}\langle\alpha|\frac{\phi'}{\sqrt{m}}|\alpha\rangle,
\quad
& \langle\alpha|\phi|\alpha\rangle = \frac{1}{\sqrt{2}}(\alpha+\alpha^*),
\nonumber\\
\langle\alpha|\Pi^2|\alpha\rangle = & -\frac{1}{2}(\alpha-\alpha^*)^2+\frac{1}{2}\langle\alpha|\frac{\phi'}{\sqrt{m}}|\alpha\rangle,
\quad & \langle\alpha|\Pi|\alpha\rangle = -i\frac{1}{\sqrt{2}}(\alpha-\alpha^*).
\label{j38a}
\end{eqnarray}
Substituting the expressions given in (\ref{j38a}) in (\ref{j34}) one finds that
\begin{eqnarray}
\left(\Delta \phi\right)^2 = \frac{1}{2}\langle\alpha|\frac{\phi'}{\sqrt{m}}|\alpha\rangle, \;\;\;
\left(\Delta \Pi\right)^2 = \frac{1}{2}\langle\alpha|\frac{\phi'}{\sqrt{m}}|\alpha\rangle.
\label{j39}
\end{eqnarray}
Hence, these coherent states minimize the uncertainity relation as
\begin{eqnarray}
\left(\Delta \phi\right)^2\left(\Delta \Pi \right)^2 = \frac{1}{4}
\langle\alpha|\frac{\phi'}{\sqrt{m}}|\alpha\rangle^2.
\label{j40}
\end{eqnarray}
The results obtained above prove that the states $|\alpha\rangle$ minimize the deformed
position-momentum uncertainty relation.

\section{\bf Examples}
\label {sec5}
Even though the algorithm given above is applicable for any exactly solvable potential,
for the sake of illustration, in the following, we consider only two examples, namely (i) linear harmonic oscillator and (ii) a new exactly solvable nonlinear oscillator which was introduced recently by Cari\~{n}ena {\it et al}\cite{car1} and construct coherent state of their associated PDMSE.  We take two examples in which the result should be known for the first example and unknown for the second.  This is because of the reason that through the first example we want to confirm the validity of our algorithm and once the results are verified we then apply the procedure to
an example whose coherent states are unknown.  Since the second example is new to the literature
we specifically consider it and construct the coherent states associated with the PDMSE.
\subsection{Harmonic Oscillator}
\label {sub sec1}
The Schr\"{o}dinger equation for the harmonic oscillator is of the form
\begin{eqnarray}
\left[-\frac {1}{2}\frac {d^2}{dy^2}+\frac{y^2}{2}\right]\psi_n(y)=E_n\psi_n(y).
\label{j41}
\end{eqnarray}
The energy values and the corresponding eigenfunctions for this problem are \cite{schiff,mathews}
\begin{eqnarray}
\psi_n(y) & = & N_n H_n(y)\exp\left[\frac{-y^2}{2}\right], \qquad N_n =\left(\frac{1}{\sqrt{\pi}2^n n!}\right)^{1/2}\\
 E_n & = & \frac{1}{2}+n, \qquad n=0,1,2,\ldots.
\label{j42}
\end{eqnarray}
respectively and $N_n$ is the normalisation constant.  Following the procedure given in Sec.2,
one can write down the exactly solvable PDMSE that share the same energy values with the linear
 harmonic oscillator is of the form
\begin{eqnarray}
\left[-\frac {1}{2}\frac {d}{dx}\left(\frac{1}{m(x)}\frac{d}{dx}\right)+ \frac{f^2}{2}+\frac{m^{''}}{8m^2}-\frac{7}{32}\frac{m'^2}{m^3}\right]\tilde{\psi}_{n}(x)
= \tilde{E_{n}}\tilde{\psi}_{n}(x),
\label{j43b}
\end{eqnarray}
where $y = f(x) = \int m^{\frac{1}{2}}dx$.

The eigenfunctions and energy values of (\ref{j43b}) can be derived from the relations
(\ref{pct7b}) and (\ref{pct6}) and read \cite{Alhai}
\begin{eqnarray}
\tilde{\psi}_{n}(x) & = & \tilde{N}_n m^{1/4} H_n(f(x)) \exp{\left[\frac{-f^2}{2}\right]},
\nonumber\\
\tilde{E}_n & = & \frac{1}{2} + n,  \qquad n=0,1,2,\ldots.
\label{j45}
\end{eqnarray}
where $\tilde{N}_n$ is the normalization constant which may or may not be equal to the constant
mass case and is being fixed by the mass distribution $m(x)$.  For the mass distributions we consider in this paper, the normalization constant $\tilde{N}_n=N_n$.

Now we construct the coherent state of the PDMSE (\ref{j43b}).  To do this let us first deduce
the ground state eigenfunction from the relation (\ref{j45}), that is
\begin{eqnarray}
\tilde{\psi}_{0}(x) =\tilde{N_0} m^{1/4}\exp\left[\frac{-f^2}{2}\right],
\label{j46}
\end{eqnarray}
with $\tilde{E}_0 = \frac{1}{2}$.  Since $\tilde{E}_0 \neq 0$ one has to subtract this ground state 
energy from (\ref{j43b}) in order to find the ladder operators which in turn fixes the 
resultant PDMSE in the factorized form (\ref{j23d}).  
Substituting the ground state solution Eq.(\ref{j46}) in Eq.(\ref{j19})
we can obtain $K$ from which one can construct the deformed potential $\phi$ by
using Eq.(\ref{j21}), that is 
\begin{eqnarray}
K=ff'-\frac{m'}{4m},\;\;\;
\phi =f.
\label{j48}
\end{eqnarray}
The annihilation and creation operators turned out to be
\begin{eqnarray}
\hat{A}=\frac{1}{\sqrt{2}}\left[m^{-1/4}\frac{d}{dx}m^{-1/4}+f\right],\;\;\;
\label{j49}
\hat{A}^\dagger = \frac{1}{\sqrt{2}}\left[-m^{-1/4} \frac{d}{dx} m^{-1/4} + f\right].
\label{j50}
\end{eqnarray}
The coherent states are obtained as
\begin{eqnarray}
|\alpha\rangle = \tilde{N_0}  m^{1/4}\exp{\left[\sqrt{2}\alpha f(x)\right]}\exp[\frac{-f^2}{2}].
\label{j51}
\end{eqnarray}

The coherent states are given for an arbitrary mass distribution.  In the following,
we consider two different kinds of mass distributions and derive an explicit form of
the coherent state.

\noindent{{\bf{Case 1}}}\\
In the nanofabrication of semiconductor devices, one observes quantum wells with very
thin layers \cite{Gos}.  The effective mass of an electron (hole) in the thin layered quantum
wells varies with the composition rate.  In such systems, the mass of the electron
may change with the composition rate which depends on the position.  As a consequence
attempts have been made to analyze such PDMS and their underlying properties for a
number of potentials and masses.  One such mass profile which is found to be useful
for studying transport properties in semiconductors is given by \cite{hetero,Gos,intertwin}
\begin{eqnarray}
m(x) = \frac{(\gamma +x^2)^2}{(1+x^2)^2},\;\;\;\gamma = constant.
\label {case11}
\end{eqnarray}
If we take the parameter $\gamma=1$, the position dependent mass is reduced to constant
mass.

We have now
\begin{eqnarray}
f(x) = \int m^{\frac{1}{2}}(x) dx = x+(\gamma-1)tan^{-1}x,
\quad -\infty <f(x)<\infty .
\label{case12}
\end{eqnarray}

The corresponding PDMSE takes the form
\begin{eqnarray}
\left[-\frac {1}{2}\frac {d}{dx}\left(\frac{(1+x^2)^2}{(\gamma +x^2)^2}\frac{d}{dx}\right)+ \frac{f^2}{2}+\frac{(\gamma-1)(3x^4+2(2-\gamma)x^2-\gamma)}{2(\gamma+x^2)^4}\right]\tilde{\psi}_n
= \tilde{E}_n\tilde{\psi}_n,
\label{j43_1}
\end{eqnarray}
where $f$ is given in (\ref{case12}).  The coherent states are given by
\begin{eqnarray}
|\alpha\rangle = N_0  \sqrt{\frac{(\gamma +x^2)}{(1+x^2)}} \exp{\left[\frac{-(x+(\gamma-1)tan^{-1}x)\left(x+(\gamma-1)tan^{-1}x-2\sqrt{2}\alpha\right)}{2}\right]},
\label{j43_2}
\end{eqnarray}
where $N_0 = (1/\sqrt{\pi})^{(1/2)}$.  In the constant mass case ($\gamma=1$) the result given
above exactly coincides with Molski (vide Eq.(17) in Ref. \onlinecite{mols}).

\noindent{{\bf{Case 2}}}\\
Another important mass profile which has been studied in graded alloys is of the form \cite{milano}
\begin{eqnarray}
m(x) = \cosh^2[\gamma x].
\label{case13}
\end{eqnarray}
so that for $\gamma=0$ one can recover the constant mass case.  A graded alloy quantum
well, typically based on $Al_xGa_{1-x}As$, will provide equispaced levels if the grading
function, that is the variation of the mole fraction $x$ along some direction is chosen to be
parabolic \cite{milano}.  Such a design of quantum well structures with some number of equispaced levels,
enable resonant interaction at all levels with monochromatic light and fully resonant
interaction.  We intend to construct the wavefunction and energy values for the PDMS with this mass
profile.

If we take the parameter $\gamma=0$, the position dependent mass is reduced to constant
mass.  In the present case we have
\begin{eqnarray}
y = f(x) = \int m^{\frac{1}{2}}(x) dx = \frac{\sinh(\gamma x)}{\gamma},
\quad -\infty <f(x)<\infty .
\label{case14}
\end{eqnarray}

The corresponding PDMSE takes the form
\begin{eqnarray}
\left[-\frac{1}{2} \frac{d}{dx} \left(\frac{1}{\cosh^2[\gamma x]}\frac{d}{dx}\right)+\frac{f^2}{2}+ \frac{\gamma^2}{16}sech^4(\gamma x) \left[7-3\cosh(2\gamma x)\right]\right]\tilde{\psi}_n = \tilde{E}_n \tilde{\psi}_n.
\label{j43}
\end{eqnarray}
where $f$ is given in (\ref{case14}).  The coherent states are given by
\begin{eqnarray}
|\alpha\rangle =N_0 \sqrt{\cosh[\gamma x]}  \exp{\left[\frac{-\sinh(\gamma x)\left(\sinh(\gamma x)-2\sqrt{2}\alpha \gamma \right)}{2\gamma^2}\right]}.
\label{j43a}
\end{eqnarray}
Here also one can check in the constant mass case ($\gamma=0$) the expression given in
(\ref{j43a}) coincides with the one given in Ref. \onlinecite{mols}.
\subsection{A Nonlinear Oscillator}
\label {sub sec2}
In a recent paper Cari\~{n}ena {\it et al} \cite{car1,car2} have considered a
non-polynomial one dimensional quantum potential representing an oscillator,
which can be considered as placed in between the harmonic oscillator
and the isotonic oscillator, and shown that it is an exactly solvable potential.
They have obtained the eigenfunctions and energies of the bound states of this potential.
Later Fellows and Smith \cite{Fel}
have shown that the potential considered by Cari\~{n}ena {\it et al}
is a supersymmetric partner potential of the harmonic oscillator.  In a recent work Kraenkel and
Senthilvelan \cite{sen}
have studied exact solutions of the PDMSE associated with this
nonlinear oscillator.  Now we construct the coherent state of the PDMSE associated
with this nonlinear oscillator.

The one dimensional potential proposed by Cari\~nena {\it et al} in \cite{car1} is of the form
\begin{eqnarray}
V(y)=\frac{1}{2}\left(y^2+\frac{8\left(2y^2-1\right)}{(1+2y^2)^2}\right).
\label{j52}
\end{eqnarray}
The wavefunction and energy values of the Schr\"{o}dinger equation corresponding to this potential
($V(y)$) is \cite{car1}
\begin{eqnarray}
\psi_n(y)=N_n \frac{{\cal{P}}_n(y)}{(1+2y^2)} \exp[\frac{-y^2}{2}], \quad n=0,3,4,\ldots
\label{j53}
\end{eqnarray}
where $N_n =\left[\displaystyle{\frac{(n-1)(n-2)}{2^n n! \sqrt{\pi}}}\right]^{1/2} $ is normalization constant and ${\cal{P}}_n(y)$ is the ${\cal{P}}$-Hermite function \cite{car1}.
The PDMSE corresponding to the potential is given by
\begin{eqnarray}
\left[-\frac {1}{2}\frac {d}{dx}\left(\frac{1}{m(x)}\frac{d}{dx}\right)+\frac{1}{2}\left[f^2+\frac{8\left(2f^2-1\right)}
{(1+2f^2)^2}\right]+\frac{m^{''}}{8m^2}-\frac{7}{4}
\left(\frac{m'}{8m^3}\right)^2\right]\tilde{\psi}_n
=\tilde{E}_n \tilde{\psi}_n.
\label{j55}
\end{eqnarray}
The wavefunctions and energy values for the PDMSE (\ref{j55})  are given by \cite{sen}
\begin{eqnarray}
\tilde{\psi}_n(x) = \tilde{N}_n m^{1/4}\frac{P_n(f(x))}{(1+2f(x)^2)} \exp[{\frac{-f(x)^2}{2}}]
\nonumber\\
 \tilde{E}_n = -\frac{3}{2} + n, \qquad n=0,3,4,\ldots.
\label{j56}
\end{eqnarray}

For the present purpose, we calculate the ground state solution from (\ref{j56}), namely
\begin{eqnarray}
\tilde{\psi}_0(x)=\frac{\tilde{N_0} m^{1/4}}{(1+2f(x)^2)} \exp[{\frac{-f(x)^2}{2}}].
\label{j57}
\end{eqnarray}
with $\tilde{E}_0 = -\frac{3}{2}$.  In this case also we have $\tilde{E}_0 \neq 0$ and so 
this ground state energy has to be subtracted from (\ref{j55}) in order 
to find the ladder operators which in turn fixes the resultant PDMSE in 
the factorized form (\ref{j23d}).  The functions $K$ and  $\phi(x)$ can be fixed of the form
\begin{eqnarray}
K=\frac{4ff'}{1+2f^2}+ff'-\frac{m'}{4m}, \\
\phi = \left[\frac{4f}{1+2f^2}+f\right].
\label{j59}
\end{eqnarray}
The abstract operators corresponding to this system are
\begin{eqnarray}
\hat{A}=\frac{1}{\sqrt{2}}\left[m^{-1/4} \frac{d}{dx} m^{-1/4} + \frac{4f}{1+2f^2}+f\right],\\
\label{j60}
\hat{A}^\dagger = \frac{1}{\sqrt{2}}\left[-m^{-1/4} \frac{d}{dx} m^{-1/4}+\frac{4f}{1+2f^2}+f\right].
\label{j61}
\end{eqnarray}
The coherent states are obtained as
\begin{eqnarray}
|\alpha\rangle = \tilde{N_0} m^{\frac{1}{4}}(x)\exp{\left[\sqrt{2}\alpha f(x)\right]}\frac{\exp[{\frac{-f(x)^2}{2}}]}{(1+2f(x)^2)}.
\label{j62}
\end{eqnarray}

Now let us consider the same two mass distributions discussed in the previous example and construct coherent states of the PDMSE (\ref{j55}).

\noindent{{\bf{Case 1}}}
Let us consider the mass profile given in (\ref{case11}).  The coherent states are found to be
of the form
\begin{eqnarray}
|\alpha\rangle = N_0 \sqrt{\frac{(\gamma +x^2)}{(1+x^2)}}  \frac{\exp{\left[-{(x+(\gamma-1)tan^{-1}x)\left(x+(\gamma-1)tan^{-1}x-2\sqrt{2}\alpha\right)}/2\right]}}
{\left(1+2\left(x+(\gamma-1)tan^{-1}x\right)^2\right)}.
\label{j43aa}
\end{eqnarray}

\noindent{{\bf{Case 2}}}
Let us consider the mass profile given in (\ref{case13}).  Repeating the procedure one
  can find the coherent states are given by
\begin{eqnarray}
|\alpha\rangle = N_0  \sqrt{\cosh[\gamma x]}\;\frac{\exp{\left[-{\sinh(\gamma x)\left(\sinh(\gamma x)-2\sqrt{2}\alpha \gamma \right)}/{2\gamma^2}\right]}}
{\left(1+2\left(\frac{\sinh(\gamma x)}{\gamma}\right)^2\right)}.
\label{j43ab}
\end{eqnarray}

One can check Eq.(\ref{j43aa}) with $\gamma=1$ and Eq.(\ref{j43ab}) with $\gamma=0$ provide the coherent
states for the potential (\ref{j52}) (constant mass) which is also unknown in the literature.

\section{\bf Perelomov's approach}
\label{sec f}

In this section, we construct coherent states by considering Perelomov's definition and show 
that the latter results agree with the ones found in this paper.
 
In Perelomov's approach, one assumes that there exists a unitary operator $D$ which acts as
a displacement operator on the ladder operators $\hat{A}$ and $\hat{A}^{\dagger}$ and $D$ be
a function of a complex parameter $\alpha$ which displaces the ladder operators according to the scheme 
\cite{Glau}
\begin{eqnarray}
D^{\dagger}(\alpha)\hat{A}D(\alpha) = \hat{A} + \alpha, \qquad D^{\dagger}(\alpha)\hat{A}^{\dagger}D(\alpha) = \hat{A}^{\dagger} + \alpha^{*}.
\label{j63}
\end{eqnarray}
Using Eqs. (\ref{j63}) and (\ref{j31a}) one can show that $D^{\dagger}|\alpha\rangle$ is
just the ground state $|0\rangle$.  Hence the coherent states can be represented as diplaced forms 
of the ground state, that is 
\begin{eqnarray}
|\alpha\rangle = D(\alpha)|0\rangle. 
\label{j64}
\end{eqnarray} 

In Sec. V, we discussed the coherent states for the harmonic oscillator and the nonlinear oscillator
(vide Eqs. (\ref{j51}) and (\ref{j62}) respectively).  Now we construct coherent states for them 
using the definition given above (Eq. (\ref{j64})) by invoking their ladder operators 
$\hat{A}$ and $\hat{A}^{\dagger}$ derived in this paper.  For the position dependent mass harmonic oscillator we have simply
$\phi = f(x)$ (vide Eq. (\ref{j48})) which in turn fixes $[\hat{A}, \hat{A}^{\dagger}] = 1$ as one expects.  
Proceeding further one finds that the displacement operator in the conventional form, that is   
\begin{eqnarray}
D(\alpha) = e^{(\alpha \hat{A}^\dagger - \alpha^{*} \hat{A})}
\label{j66}
\end{eqnarray}
which acting on the ground state yields coherent states as combinations of wavefunctions, 
$\tilde{\psi}_n(x)$ (vide Eq. (\ref{j45})), as
\begin{eqnarray}
|\alpha\rangle = e^{\frac{-|\alpha|^2}{2}}\sum^{\infty}_{n=0} \frac{\alpha^n}{\sqrt{n!}}\tilde{\psi}_n(x). 
\label{j67}
\end{eqnarray}
We arrived an expression which is same as that of constant mass Schr\"{o}dinger equation \cite{How}. 

To discuss the general case, we recall here that the ladder operators $\hat{A}$ and $\hat{A}^{\dagger}$
(vide Eqs. (\ref{j22}) and (\ref{j23})) and the Hamiltonian $H$ (Eq. (\ref{j23a})) satisfy the
relations  
\begin{eqnarray}
[\hat{A},  \hat{A}^{\dagger}]= \frac{\phi'}{\sqrt{m(x)}}, \quad [H, \hat{A}]= -\frac{\phi'}{\sqrt{m(x)}} \hat{A}, 
\qquad [H, \hat{A}^{\dagger}]= \hat{A}^{\dagger}\frac{\phi'}{\sqrt{m(x)}}. 
\label{j68}
\end{eqnarray}
To construct a displacement operator for a general potential (which evolve with its $\hat{A}$ and $\hat{A}^{\dagger}$ 
as per the relations in (\ref{j68})) one may consider $D(\alpha)$ of the form $D(\alpha) = e^{ih(\alpha)}$
and try to fix an appropriate expression for $h(\alpha)$. Imposing the restriction that $D(\alpha)$ be an unitary one 
and should satisfy the requirements given in (\ref{j63}) we find that
\begin{eqnarray}
[h(\alpha), \hat{A}] = i\alpha, \quad[h(\alpha), \hat{A}^{\dagger}] = i\alpha^{*}.
\label{j69}
\end{eqnarray}
A compatiable solution which satisfies both the equations in (\ref{j69})
can be found as $h(\alpha) = -i \sqrt{2} \alpha f(x)$ which in turn
fixes coherent states of the form
\begin{eqnarray}
|\alpha\rangle = e^{\sqrt{2}\alpha f(x)}|0\rangle,
\label{j70}
\end{eqnarray}
where $\alpha$ is purely imaginary one.  Eq. (\ref{j70}) also coincides with  Eq. (\ref{j31a}). 
 
To confirm the validity of (\ref{j70}) one can again consider the position dependent mass harmonic oscillator
with $f(x) = \frac{1}{\sqrt{2}}(\hat{A}+\hat{A^{\dagger}})$ (vide  Eqs. (\ref{j37}) and (\ref{j48})).  In this case
Eq. (\ref{j70}) gives an expression displayed in (\ref{j67}). 
\section{Conclussion}
\label{sec 6}
In this paper, by extending the ideas given in Ref. \onlinecite{mols}, we have developed an algorithm
to construct coherent states for the exactly solvable one dimensional PDMSE.  The algorithm essentially
consists of two steps: (i) to obtain ground state eigen function of the PDMSE and (ii) to build
suitable creation and annihilation operators.  We have utilized point canonical transformation
method which is being widely used in the literature and obtained ground state eigen function
of the PDMSE.  In the second step we have used the deformed ladder operators and obtained
explicit expression for the deformed superpotential in terms of mass distribution $m(x)$ and
its derivative.  We have also proved that the states minimize the generalized position-momentum
uncertainty relation.  Through this way we have established a general algorithm to construct
coherent states for an exactly solvable PDMSE.  Even though our algorithm is valid for any exactly solvable
potential, for the illustration purpose, we have considered only two examples and demonstrated the
underlying ideas.  In this process we have included an example (Example 2) whose coherent states are
new to the literature.  We have also considered Perelomov's approach and constructed coherent states
for PDMSE.  The coherent states obtained through both the procedures agree with each other.
\section*{Acknowledgements}
MS wishes to thank Department of Science and Technology(DST), Government of India and DST-CNPq
for the financial support through major research projects.

\section*{References}

\end{document}